# Early detection of the advanced persistent threat attack using performance analysis of deep learning

Javad Hassannataj Joloudari[1,2], Mojtaba Haderbadi[1], Amir Mashmool[1], Mohammad GhasemiGol[1], Shahab S.[3,4], Amir Mosavi[5,6,7,8]

[1] Computer Engineering Department, Faculty of Engineering, University of Birjand, Birjand, Iran
[2] Department of Information Technology, Mazandaran University of Science and Technology, Babol, Iran
[3] Institute of Research and Development, Duy Tan University, Da Nang 550000, Vietnam
[4] Future Technology Research Center, College of Future, National Yunlin University of Science and Technology 123, University Ro ad, Section 3, Douliou, Yunlin 64002, Taiwan
[5] Institute of Automation, Obuda University, 1034 Budapest, Hungary
[6] School of Economics and Business, Norwegian University of Life Sciences, 1430 Ås, Norway
[7] Faculty of Civil Engineering, Technische Universität Dresden, 01069 Dresden, Germany
[8] Department of Informatics, J. Selye University, 94501 Komarno, Slovakia

**Abstract**

One of the most common and important destructive attacks on the victim system is Advanced Persistent Threat (APT)-attack. The APT attacker can achieve his hostile goals by obtaining information and gaining financial benefits regarding the infrastructure of a network. One of the solutions to detect a secret APT attack is using network traffic. Due to the nature of the APT attack in terms of being on the network for a long time and the fact that the network may crash because of high traffic, it is difficult to detect this type of attack. Hence, in this study, machine learning methods such as C5.0 decision tree, Bayesian network and deep neural network are used for timely detection and classification of APT-attacks on the NSL-KDD dataset. Moreover, 10-fold cross validation method is used to experiment these models. As a result, the accuracy (ACC) of the C5.0 decision tree, Bayesian network and 6-layer deep learning models is obtained as 95.64%, 88.37% and 98.85%, respectively, and also, in terms of the important criterion of the false positive rate (FPR), the FPR value for the C5.0 decision tree, Bayesian network and 6-layer deep learning models is obtained as 2.56, 10.47 and 1.13, respectively. Other criterions such as sensitivity, specificity, accuracy, false negative rate and F-measure are also investigated for the models, and the experimental results show that the deep learning model with automatic multi-layered extraction of features has the best performance for timely detection of an APT-attack comparing to other classification models.

## 1. Introduction

Providing information security is one of the main problems of the companies and organizations, and they constantly try to ensure that their data and information are not compromised due to the accidents and attacks [1]. The attacks and activities of the attackers have become more complicated and targeted owing to the progress and growth of the cyberspace.

The Authors' Email: javad.hassannataj@birjand.ac.ir (Javad Hassannataj Joloudari), mojtabah@birjand.ac.ir (Mojtaba Haderbadi), amir.mashmool@birjand.ac.ir (Amir Mashmool), shamshirbands@yuntech.edu.tw (Shahab S.), amir.mosavi@kvk.uni-obuda.hu (Amir Mosavi)

According to Gartner, budgets have risen from $114 billion in 2018 to more than $124 billion in 2019. Information technology security leaders in companies agree to a 72% budget increase in 2020 to take steps such as continuous staff training, awareness and skill enhancement and reduce the damage caused by intrusion into their systems. Today, most of the attacks that threaten companies are targeted and long time, some of which are known as Advanced Persistent Threats (APT) [2]. The term APT was first introduced in 2006 by US Army Air Force specialists regarding unknown intrusion activities [3]. APT attacks are carried out by a group of well-funded attackers with a pre-determined plan to gain access to the confidential information or data of the companies. This attack is a multi-step and persistent attack through which the attacker can remain in the victim system for several months with full awareness [4]-[7].

An APT attack has three characteristics [3], [4], which include 1) threats; the ability of the attacker to access confidential information, 2) advanced; using advanced techniques to complete the attack cycle by the attacker, and 3) persistent; the slow process of the attacker to reach the defined goal. Consequently, an APT attack can be favorable for the attacker from three points of view. The first is that the attacker has unlimited time to attack. Second, the attacker can seize unlimited resources, and third, the organizations need to focus on their business strategies rather than spending all of their resources on defensive strategies [5].

Examples of APT attacks that have occurred in recent years are listed below.

EPIC TURLA, which was identified by Kaspersky, aimed to infect the systems of government agencies, state departments, military agencies and embassies in more than 40 countries worldwide [6].

Deep panda was an attack carried out to obtain the information of the staff of the US Intelligence Service, and was probably of Chinese origin. The attackers used the Deep panda code to endanger the information of more than 4 million employees [7].

In addition, a group from Russia known as Fancy Bear, Pawn Storm and Sednit was identified by Trend Micro in 2014 that launched attacks on military and government targets in Ukraine, Georgia, NATO and US defense allies [5].

In an APT attack, attackers use various methods to intrude, two of which are mentioned as follows: I) Zero Day; Attackers identify the weaknesses of a company or an organization and use them to damage the systems [4], II) Targeted phishing; In this method, attackers use infected emails that contain malware to intrude) the systems [4]. In APT attack, attackers try to find the codes of the target systems and programs at the beginning of the task to intrude the systems. As attacks become more complicated, traditional security systems such as firewalls, web and email protectors and scanners are no longer suitable for defending and preventing damages. One of the serious issues and challenges associated with APT attack is lack of a high-precision and real-time detection system. The other challenges are that the attacker is able to invisibly analyze the victim system using structured models and can be placed in the target system for a long time [2], [8].

Therefore, the methods used by researchers to detect APT attacks are as follows.

- Detection models based on machine learning algorithms, including linear support vector machine, Quadratic SVM, Cubic SVM, Fine Gaussian SVM, Medium Gaussian SVM, Coarse Gaussian SVM [8] as a subset of SVM methods as well as Complex tree, Medium tree and Simple tree for decision tree [9].
- Detection models based on mathematical models, such as hidden Markov model [10].
- Methods and approaches for automatic extraction of features using attack graph [11].
- Techniques to reduce false detection, such as Duqu tool [12].
- Detection of all attack steps using tools, such as SpuNge [13].

Although aforementioned schemes can be relatively appropriate detection methods for dealing with APT attacks, they cannot perform timely detection when attacks occur in real-time. In addition, these methods have high false negative and positive rates, which are important criteria that can indicate the effectiveness of a method in correct detection of the attack. None of these methods provide a system model capable of detecting a new attack pattern, high generalizability and high flexibility. Finally, lack of proper process on the dataset of the attacks in these methods is quite obvious. Consequently, we decided to examine and investigate machine learning methods such as C5.0 decision tree, Bayesian network and deep neural network on the NSL-KDD dataset. As a result, it can be stated that deep neural network method greatly reduces the weaknesses of the mentioned methods and can be a powerful approach to detect an APT attack on the considered dataset, since deep learning model provides high detection accuracy (ACC) as well as automatic extraction of the main features of the attack. In other words, according to our latest and greatest knowledge about APT attack detection, we can reasonably argue that among the available methods, deep learning method [14] as an intelligent method that is used today for large data sets in different organizations, provides the best performance.

It is noteworthy that it is the first time that C5.0 decision tree, Bayesian Network and deep learning models are utilized to detect APT attack on the NSL-KDD dataset. In this paper, deep learning model as a proposed model along with Bayesian Network and C5.0 decision tree models is implemented on the relatively large NSL-KDD dataset. In brief, the contributes of the article are as follows:

- Improving the detection accuracy by analyzing the data through deep learning model in comparison with C5.0 decision tree and Bayesian classification models.
- Improving deep neural network training by testing the existing data through Maxout method and cross-validation in order to avoid over-fitting and increase generalizability.
- Proposing a 6-layer deep learning model by automatic extracting and selecting the features in the hidden layers of the neural network.

The remaining sections of this paper are organized as follows. We explain related work in Section 2. Section 3 describes our proposed methodology regarding APT attack detection using classification models. The evaluation of the models' performance is accomplished and analyzed in section 4. Section 5 presents the experimental results. Section 6 represents "Results and Discussions". Finally, we conclude our paper with some suggestion for future research works in Section 7.

## 2. Literature Review

The detection methods of APT attacks that have been introduced up to now, have disadvantages, such as high rate of false detection of the attacks and lack of real-time detection. Due to the fact that APT attack uses secret and intelligent techniques, and can stay in the system for months, therefore, traditional intrusion detection systems cannot detect these attacks, because they are usually based on pattern or signature and use applications to detect APT [15]-[17].

The APT attack detection methods with different criteria that have been studied by researchers so far have been investigated in the following. Among these methods that have led to better detection of the attacks are machine learning-based methods.

Salama et al. have used Deep Belief Network (DBN) method combined with Support Vector Machine (SVM) to detect intrusion into the NSL-KDD dataset [18]. The combined DBN-SVM method through 40% of the training dataset with 92.84% detection accuracy provides better performance in comparison to the SVM and DBN methods.

Despite the growth and spread of APT attacks, no specific research and study has been conducted about this attack, and most of the investigations about APT consist the attack patterns and its general information, and automatic detection methods have not been taken into consideration [19]-[22]. One of the most vulnerable platforms is mobile phones, which are very popular for attackers [23].

Aziz et al. have used classification algorithms including Naive Bayes (NB), Multilayer Perceptron Neural Network and Decision trees in order to detect Denial-of-Service (DoS), User to Root (U2R), Remote to local (R2L) and Probe attacks on the NSL-KDD dataset [24]. Their results show that the Naive Bayes classification method has better detection accuracy for R2L and U2R attacks, so that for R2L attack, 32.90% and 21.07% accuracies were obtained using all the training data and 20% of the training data, respectively, and also, for U2R attack, 20.35% and 16.28% accuracies were obtained using all the training data and 20% of the training data, respectively. Then, they achieved high accuracy of up to 82% for DoS attack and 65.4% accuracy for Probe attack using J48 decision tree.

In some cases, the introduced methods and tools are not complete, meaning that they may only be able to detect the vulnerability of the environment or network and not be able to detect the attack in the environment, as in [25] that Johnson and Hogan have proposed a method to investigate whether the network environment is vulnerable to an APT attack or not. This tool allows the network security administrators to check the vulnerability of the network environment after initial configuration and then make changes if necessary.

Ingre and Yadav have utilized Artificial Neural Network (ANN) method to detect intrusion into the NSL-KDD dataset [26]. Their results indicate that the ANN method reaching 81.2% and 79.9% detection accuracies for binary class (i.e. normal and attack statuses) and 5 classes (i.e. DoS, Probe, R2L, U2R attacks and normal status), respectively, provides better performance comparing to Self-Organization Map (SOM) method with a detection accuracy of 75.49%.

Friedberg et al. state that in order to detect an APT attack, the attack detection system requires a large amount of information and input data [27]. In addition, at the end of the detection process, what the detection system presents as a result is highly complex, and it is very difficult for security analysts to understand it.

Guo et al. have proposed a two-layer combined approach to detect intrusion into the KDD-99 dataset and the Kyoto University Benchmark Dataset (KUBD) [28], which consists of two anomaly detection components and one misuse detection component. In the first step, an Anomaly Detection method Based on the change of the Cluster Centres (ADBCC) has been used to construct the first component of the anomaly detection, where the cluster centres are obtained using the K-means algorithm. In the second step, by applying the K-nearest neighbor (k-NN) algorithm, two different detection components, including anomaly detection and misuse detection have been constructed. In fact, the anomaly detection component created in the first step participates in the construction of the two detection components in the second step. Their results in terms of the detection accuracy indicate that the proposed combined approach with 93.29% accuracy has better performance comparing to ADBCC method with 92.71% accuracy for both normal and attack classes on the KDD-99 dataset. Furthermore, the proposed combined approach with 95.76% accuracy performed better than the ADBCC method with 92.85% accuracy on the KUBD dataset.

Mazraeh et al. have used machine learning algorithms such as SVM, Naive Bayes and J48 decision tree as well as classification to detect intrusion into the KDD-99 dataset [29]. The feature selection method they employed was Information Gain. Their experiments show that the J48 decision tree method in combination with the AdaBoost method has the highest accuracy of 97% compared to other methods.

Bhatt et al. have proposed a method to predict and detect APT attacks [30]. Due to the fact that this attack is dynamic and can be developed in several directions in parallel, a combination of attack and defense patterns have been used in this model. To implement the procedure, Apache Hadoop has been performed with a logical layer that includes Information Gathering, Weaponization, Delivery, Exploitation, Installation, Command and Control (C2) and Actions steps, and is capable of predicting and detecting APT attacks. Each of these steps is necessary to pursue the goals.

In an APT attack, since the attacker is following the program with great planning and precision, he makes every effort to behave normally on the network so that the detection tools do not notice his presence, and it makes it difficult to detect the attack. However, in [31], Marchetti et al. have proposed a method to detect the infected hosts. The method receives the network traffic and displays a list of infected hosts at the end of the process.

Raman et al. have utilized an intrusion detection technique using a Hyper Graph based Genetic Algorithm (HG-GA) for parameter setting and feature selection in the Support Vector Machine (SVM) on the NSL-KDD dataset [32]. The proposed HG-GA SVM method with 96.72% accuracy performs better than the Grid-SVM, PSO-SVM, GA-SVM, Random Forest and Bayes Net methods.

Bodström and Hämäläinen have utilized Observe-Orient- Decide-Act (OODA) loop and Black Swan Theory for detection and identification of APT attacks [33]. In this paper, without manipulating and reducing the features, the network data stream is transferred to the detection process. It has been suggested that in order to better detect an attack, the most important factor in the attack must be identified, which in the case of APT, is communication factor, and in result, the network stream must be recorded to identify the attack.

Ghafir et al., by receiving the network traffic and after analyzing the data, have implemented algorithms such as decision tree, various SVM models, Nearest neighborhood and Ensemble on the data [8]. They have observed that the SVM linear algorithm has the best result with 84.8% accuracy. Finally, they have introduced a system called machine learning-based system (MLAPT). It is necessary to mention that they have calculated only the accuracy parameter for the algorithms.

Chu et al. have used the NSL-KDD database to detect the attack and have utilized the PCA method to decrease the size of the classified dataset [9]. They have concluded that the SVM algorithm with the radial basis function (RBF) as the kernel has better performance comparing to the classification algorithms such as multilayer perceptron (MLP), decision tree of J48 and Naive Bayes reaching a detection accuracy of 97.22%.

Bodström and Hämäläinen have introduced a model based on a theoretical approach or idea regarding APT attacks, and have stated that the APT attack is a persistent and multi-step attack that uses the entire network stream as input [14]. As a result, experiments demonstrate that the deep learning stack that utilizes sequential neural networks achieves a better and more flexible architecture for the APT attack detection.

Statistical methods have also been used to detect APT attacks. Hidden Markov model is one of these methods. In [10], Ghafir et al. have developed a system that can be effective in both predicting and detecting APT attacks. The system consists of two parts or sections, the first of which examines the correlation of the warnings, and the second part uses the Markov model to decrypt the attack, and the count of warnings or steps of the APT attack is considered to be 4, and the system can estimate the sequence of attack steps with an accuracy of 91.80%.

## 3. Methodology

In this study, we have used RapidMiner[1] simulator for the APT attack detection and classification process. The methodological process is illustrated in Figure 1.

According to Figure 1, the proposed methodology includes 7 modules, each of which will be described in detail in the following. In this study, the modules include data collection from an external source, pre-processing, segmentation, classifiers, model evaluation criteria, selection of the best model and extraction of the features.

### 3.1. Used Dataset

We used the NSL-KDD dataset [9], [34], to detect APT attacks. This dataset includes 148517 data samples and includes 125973 training dataset and 22544 test dataset.

The values distribution of the NSL-KDD dataset for all the attack classes is presented in Table 1.

**Table 1.** Values Distribution of the NSL-KDD dataset for all the attack classes.

| Training dataset | | Testing dataset | |
|---|---|---|---|
| **Class** | **values** | **Class** | **Values** |
| Normal | 67343 | Normal | 9711 |
| DoS | 45927 | DoS | 7458 |
| R2L | 995 | R2L | 2754 |
| U2R | 52 | U2R | 200 |
| Probe | 11656 | Probe | 2421 |
| Sum | 125973 | Sum | 22544 |

According to Table 1, the NSL-KDD dataset includes 41 features. In addition, a description of the NSL-KDD dataset with corresponding features is given in Table 2.

---

[1]https://docs.rapidminer.com/latest/studio/operators/modeling/predictive/neural_nets/deep_learning.html

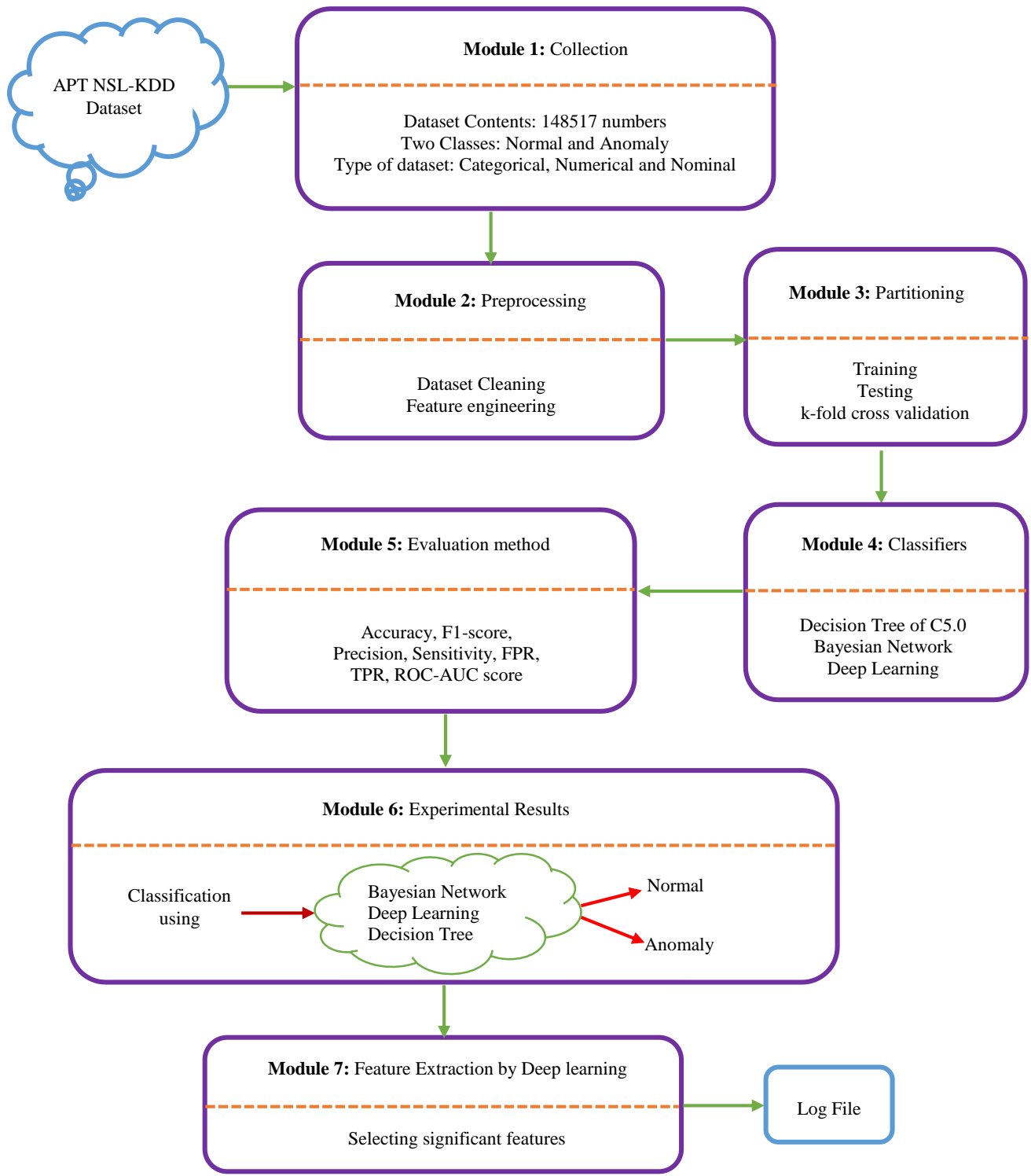

**Figure 1.** Proposed model.

**Table 2.** Features of the NSL-KDD dataset [34], [35].

| No. | Feature Name | Data Type | No. | Feature Name | Data Type |
|---|---|---|---|---|---|
| 1 | Duration | Numerical | 22 | Is_guest_login | Categorical |
| 2 | Protocol_type | Nominal | 23 | Count | Numerical |
| 3 | Service | Nominal | 24 | Srv_count | Numerical |
| 4 | Flag | Nominal | 25 | Serror_rate | Numerical |
| 5 | Src_bytes | Numerical | 26 | Srv_serror_rate | Numerical |
| 6 | Dst_bytes | Numerical | 27 | Rerror_rate | Numerical |
| 7 | Land | Categorical | 28 | Srv_rerror_rate | Numerical |
| 8 | Wrong_ fragment | Numerical | 29 | Same_srv_rate | Numerical |
| 9 | Urgent | Numerical | 30 | Diff_srv_rate | Numerical |
| 10 | Hot | Numerical | 31 | Srv_diff_host_rate | Numerical |
| 11 | Num_failed logins | Numerical | 32 | Dst_host_count | Numerical |
| 12 | Logged_in | Categorical | 33 | Dst_host_srv_count | Numerical |
| 13 | Num_compromised | Numerical | 34 | Dst_host_same | Numerical |
| 14 | Root_shell | Categorical | 35 | Dst_host_diff_srv_rate | Numerical |
| 15 | Su_attempt ed | Categorical | 36 | Dst_host_same_src_port_rate | Numerical |
| 16 | Num_root | Numerical | 37 | Dst_host_srv_ | Numerical |
| 17 | Num_file_creations | Numerical | 38 | Dst_host_serror_rate | Numerical |
| 18 | Num_shells | Numerical | 39 | Dst_host_srv_serror_rate | Numerical |
| 19 | Num_access_files | Numerical | 40 | Dst_host_rerror_rate | Numerical |
| 20 | Num_outbo und_cmds | Numerical | 41 | Dst_host_srv_rerror_rate | Numerical |
| 21 | Is_hot_login | Categorical | | | |

### 3.2. *Pre-processing*

In this paper, machine learning and deep learning approaches are used. One of the steps in these approaches is pre-processing of the data, which necessitates analyzing the data. In the pre-processing module on the NSL-KDD dataset, the data needs to be usable and performable for the classifiers in the next modules. Therefore, for this module, we have investigated the missing data after extracting the samples using RapidMiner software and we have found out that there is no missed and unvalued data. Additionally, we have not used feature selection in this module, since feature selection methods cannot have great effects on analyzing the NSL-KDD dataset here. It should be noted that we will investigate and execute feature extraction as an important module in the seventh module. However, feature engineering [36] is utilized in this subsection, i.e. we first need to specify the features of the APT attack in the relevant dataset in terms of their type. According to the features of the NSL-KDD dataset [35], the data types are as categorical, numerical and nominal data.

In this study, 4 attack classes including DoS, R2L, U2R and Probe are grouped under one class called anomaly as well as normal status as normal class. Finally, the dataset used in this paper is changed into two classes of anomaly and normal.

Furthermore, in order to complete the pre-processing of the data, we have aggregated and integrated the training and testing data sets so that the count of normal and anomaly samples is 77054 and 71463, respectively.

### 3.3. *Partition of the NSL-KDD dataset*

As mentioned in subsection 3.1., the NSL-KDD dataset used in this research includes 148517 samples, and as data segmentation, we consider 90% of them for training and 10% for testing. In addition, we have used 10-fold cross validation method, and for examining the proposed models, 0.9 of the data is used for training and the remaining 0.1 of the data is used for testing for each fold. In section 5, giving the experiment results, the data classification will be explained in more detail.

Note that in this study for the NSL-KDD dataset, the attack classes including DoS, R2L, U2R and Probe are grouped under a class called anomaly and also normal status as normal class.

### 3.4. *Classification models*

#### *3.4.1. C5.0 Decision tree*

In the process of improving decision tree models, C5.0 decision tree is the latest generation of the decision tree models, including CHAID, ID3 and C4.5 [37], [38]. The main task of the decision tree is to create rules that can help the security experts to detect the type of the input data based on the constructed model. A decision tree model consists of a number of nodes and branches so that the leaves (external nodes) represent normal and anomalous classes or a set of answers, and in other nodes (internal nodes), the decisions are made based on one or more features. A decision tree diagram with a depth of 5 is shown in Figure 3.

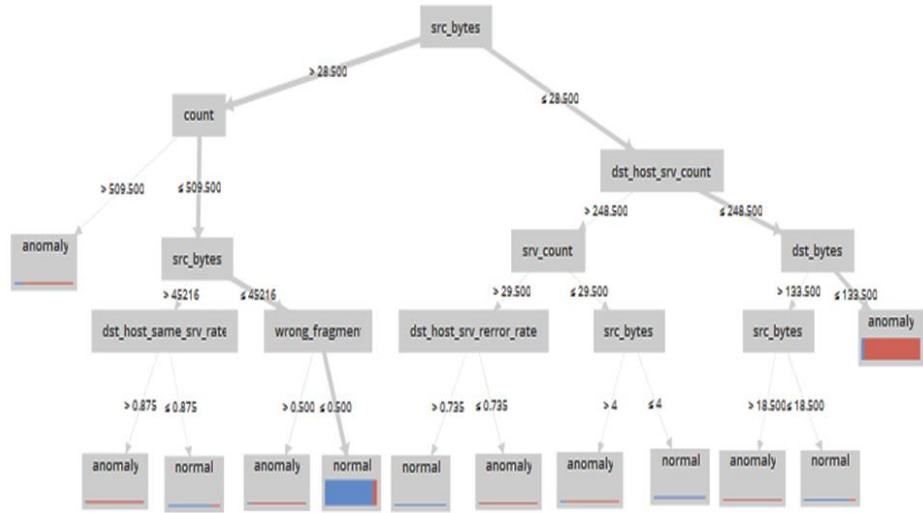

**Figure 3.** C5.0 decision tree diagram on the NSL-KDD dataset.

An important preference of the C5.0 decision tree model for testing features is the gain ratio. A higher gain ratio indicates a better model [37], [39]. The gain ratio is calculated as below:

$$GainRatio(K,C) = (Gain(K,C)/SplitInfo(K,C))  \qquad (1)$$

According to *Equation 1*, $SplitInfo(K,C)$ and $Gain(K,C)$ are calculated as follows:

$$SplitInfo(K,C) = Info_{Entropy}(|C_1|/|C|,...,|C_i|/|C|) \qquad (2)$$
$$Gain(K,C) = Info_{Entropy}(K) - Info_{Gain}(K,C) \qquad (3)$$

where $K$ is the count of the features and $C_i$ is the partition of the $C$ derived by the value of $K$.

Thus, are formulated as follows: $Info_{Gain}(K,C)$ and $Info_{Entropy}(K)$

$$Info_{Entropy}(K) = -\sum_{i=1}^{N \in C_i} P_i \log_2 P_i \qquad (4)$$

$$Info_{Gain}(K,C) = -\sum_{i=1}^{N \in C_i} P_i \times Info_{Entropy}(K_i) \qquad (5)$$

According to *Equations 4* and *5*, $P$ is calculated as follows:

$$P = (|C_1|/|S|, |C_2|/|S|,...,|C_i|/|S|) \qquad (6)$$

where $|S|$ is the count of examples in set of $S$ and $P$ is the probability distribution of partition ($C1, C2, …, Ci$).

### 3.4.2. Bayesian network model

Another classification model in data mining is Bayesian network classification model. The philosophy of this model is based on a possible framework for solving classification problems. According to Bayes' theorem, the classification of events is formed based on the probability of occurring or not occurring an event so that the probability of an event is calculated and classified [39]. In the Bayes' theorem, we have the following probabilities:

$$P(D|B) = P(B,D)/P(B) \qquad (7)$$
$$P(D|B) = P(B,D)P(D)/P(B) \qquad (8)$$

In fact, the Bayesian Network model has a graphical scheme that represents prediction variables and their eventual connections using a directed or non-circular signal graph. The nodes are also a prediction variable in the graph [39], [40].

*3.4.3. Deep learning model*

The philosophy of deep learning is derived from the architecture of biological neural networks in human brain under artificial neural networks, which is a branch of machine learning and artificial intelligence. In nerve cells and neurons, information and data are in the form of pulses or electrical signals that enter and leave the cell. In other words, nerve cells decode through tagging and assigning features and items to different categories and classes so that a series of changes and processing are performed on the cell nucleus. These changes and processes are learned during human life, and the so-called neural network structure is trained during human life. A similar process is seen in deep learning [41]-[48]. In deep learning, we deal with multi-layered deep neural networks, which introduce multi-layered learning of the features as the main characteristic. These layers are called hidden layers in the neural network, and a network is considered as a deep learning network, when it includes more than two hidden layers. In general, this model has 3 types of layers:

- Input layer: Receives input data related to features.
- Hidden layer: Data patterns are extracted in this layer.
- Output layer: Data processing results are related to this layer.

It is necessary to mention that the advantage of a deep neural network is having lots of hidden layers, which makes it different from superficial artificial neural network that has a single hidden layer. This means that deep neural network is able to do more complex tasks. The structure of a deep network is such that the data is transferred from one hidden layer to another so that simpler features are recombined and recomposed as complex features.

For example, consider a two-layered neural network in which a three-dimensional input can be connected to four other neurons of different weights in one layer. This process is similar to feature extraction process, i.e. an input from a 3-dimensional space is mapped to a new 4-dimensional space, which can be known as the feature space. In other words, the inputs are transformed to a series of features that are good and useful features. In machine learning, after the feature extraction process, we have algorithm learning process, i.e. the features are used as inputs to a classification algorithm, which learns to detect the class of the inputs. We had the same rules and principles for these methods in Subsections 3.4.1 and 3.4.2.

In this two-layer neural network, there is a layer called the feature layer or hidden layer, the outputs of which are the feature space and also the inputs to the last layer. The last layer is called the classification layer, which specifies the class of the input data related to the features. The two-layer neural network is shown in Figure 4 adopted from [39], [42].

According to Figure 4, the output generation process is such that if we multiply each of the input dimensions by a coefficient or so-called weight, and then pass sum of them through a nonlinear function, a new output is generated.

This is similar to the process that we have in a nerve cell, meaning that the input changes during the passage through the cell. These changes are weights, and the nonlinear function results in new outputs. The set of inputs, weights and nonlinear function are called a layer in artificial neural networks, and this layer must be well trained. Neural network training means finding weights to transform inputs into expected outputs. Therefore, it's the weights that are trained, and the nonlinear function is usually added to increase the network capability.

As mentioned before, the deep learning model of a neural network includes more than two middle or hidden layers.

For example, in Figure 5, a deep network with 3 hidden layers [49] using RapidMiner tool is illustrated.

In a 3-layer network, low-level, middle-level and high-level or much more complex features are extracted in the first, second and third layers, respectively. At the output of this network, we have the classification of the input data that specifies the class type of the data.

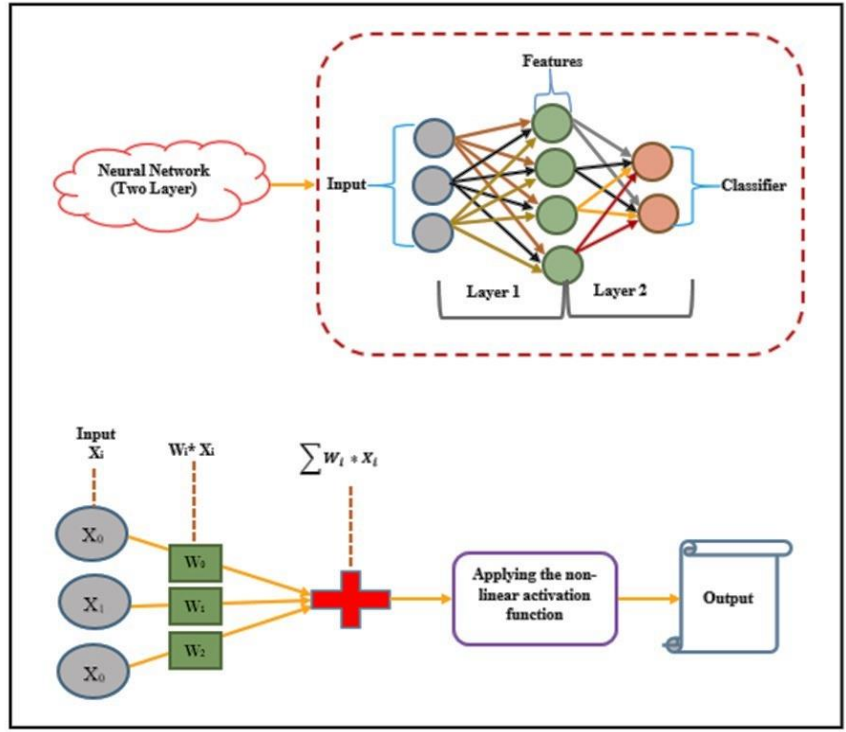

**Figure 4.** Multi-layered neural network.

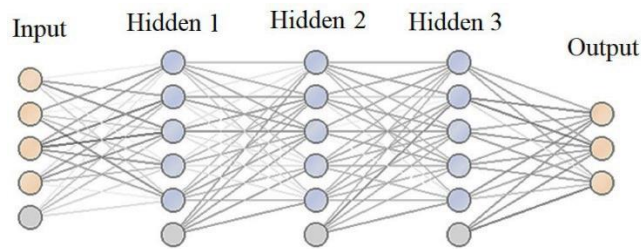

**Figure 5.** Neural network with 3 hidden layers.

Hence, the goal of deep learning is to discover several levels of distributed representations of the input data so that by creating features in the lower layers, it can differentiate the factors of changes in the input data and then combine these representations in the higher layers [50].

In addition, one of the noticeable advantages of a deep network is that deep learning model performs very well on unstructured data and has a higher accuracy comparing to machine learning models such as decision tree, Bayesian network, support vector machine, etc., but in practice, requires a large amount of training data along with appropriate hardware and software. Furthermore, one of the most important capabilities of a deep learning network is the ability to extract features automatically. Deep neural network also has a high generalizability, meaning that in addition to the data being trained, if the network receives new data that is similar to the training data, it can detect the data with high accuracy, which is called high generalization ability. In this paper, a 6-layer deep learning model with 4 hidden layers sized 50x50 in the 10 epoch range by 10-fold cross validation method is used.

Additionally, the nonlinear activation function used, which determines the activity of neurons in the hidden layers of the network, is determined by Maxout [51]. The Maxout function selects the maximum coordinates for the network input vector, and is utilized in this research to avoid data over-fitting and improve network training. The Sofmax function is also used to classify the output layer. The deep learning model used in this paper is performed in RapidMiner software. The proposed deep learning model is shown in Figure. 6.

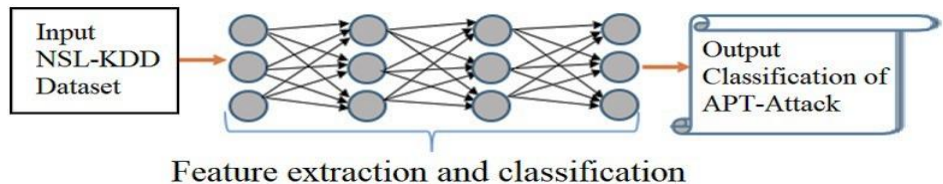

**Figure 6.** Proposed deep learning model.

According to Figure 6, after entering the data into the deep network, the extraction of features and classification of attacks is performed in combination and simultaneously, and no other method is required to extract the features, because feature extraction is performed automatically in deep network. Finally, the attack classification is accomplished after applying the nonlinear function.

## 4. Method Evaluation

In this paper, the confusion matrix is used to evaluate the proposed models [37]-[39], [52]. This matrix includes 4 elements, including True Positive (TP), False Positive (FP), True Negative (TN) and False Negative (FN). The basic definitions of these 4 elements are as follows.
- TP: Represents that when an alert is generated, then an APT attack occurs.
- FP: Represents that when an alert is generated, but an APT attack does not occur.
- TN: Represents that when an alarm is not generated, then an APT attack does not occur.
- FN: Represents that when an alert is not generated, but an APT attack occurs.

The confusion matrix is shown in Table 3.

**Table 3.** Confusion matrix for detection of APT attack.

| The Actual class | The predicted class | |
|---|---|---|
| | **Anomaly** | **Normal** |
| Positive | True Positive | False Positive |
| Negative | False Negative | True Negative |

Consequently, according to the confusion matrix, we have used 7 criteria to evaluate three models including Bayesian, C5.0 decision tree and deep learning. The criteria are accuracy, F-measure or F1-score, precision or positive predictive value (PPV), specificity (TNR), sensitivity or true positive rate (TPR), FPR and ROC-AUC score [39].

These criteria are formulated based on the following equations:

$$TNR = TN/TN + FP \tag{9}$$

$$TPR = TP/TP + FN \tag{10}$$

$$Accuracy = TP + TN/TP + TN + FP + FN \tag{11}$$

$$precision = TP/TP + FP \tag{12}$$

$$recall = TP/TP + FN \tag{13}$$

$$F - measure = 2 * \frac{precision * recall}{precision + recall} \tag{14}$$

Furthermore, FPR and FNR criteria show the type of false, and FPR is a more important criterion than FNR in terms of false determination and effectiveness. These criteria are formulated as follows:

$$FPR = 1 - TNR \tag{15}$$

$$FNR = 1 - TPR \tag{16}$$

These criteria are calculated through 10-fold cross validation method. The results of the classification models will be analyzed in the next Section.

## 5. Experimental Results

The results of the Bayesian, C5.0 decision tree and deep learning classification models are investigated and analyzed in this Section. Since the purpose of this article is detection and classification of APT attacks on network, we have used the NSL-KDD dataset to detect the APT attacks. In the detection process, after receiving the data and pre-processing, we have used classification models such as Bayesian, C5.0 decision tree and 6-layer deep learning. In addition, to evaluate the models in the output, criteria such as accuracy, precision, false positive rate (FPR), false negative rate (FNR), sensitivity, specificity and F-measure have been extracted as experiment results.

In this experiment, 10-fold cross validation method has been utilized to classify the dataset so that in each fold, 0.9 of the data has been used for training and the remaining 0.1 of the data has been used to test the performance of the proposed models, and this process has been repeated 10 times.

Moreover, in order to evaluate the generated models, control the training process, prevent over-fit of the data and improve the generalization, we have considered 90% of the training data, 80% of which is for training and 20% is for validation of the models in 10 epochs. Figure 7 illustrates the training, testing and validation process of the proposed models.

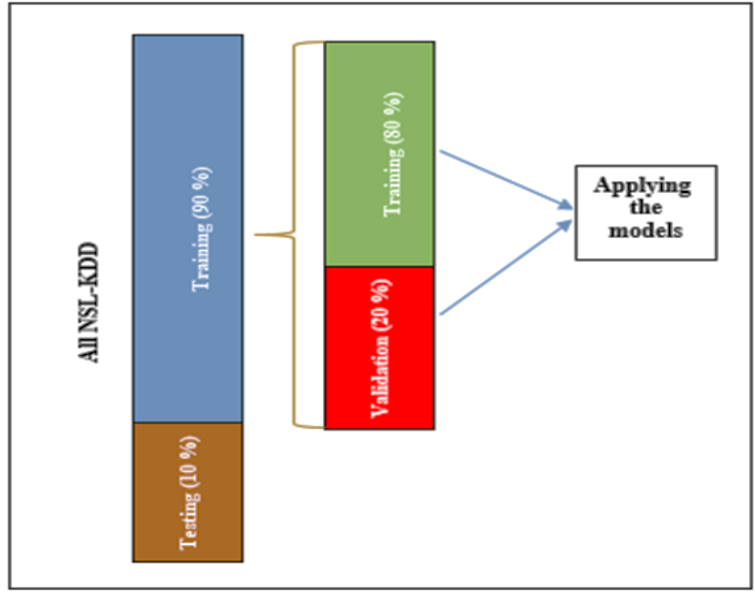

**Figure 7.** Training, testing and validation process of the proposed models.

The results based on the evaluation criteria are given in Table 4 for the Bayesian, C5.0 decision tree and deep learning classification models.

According to Table 4, the accuracy of the Bayesian network, C5.0 decision tree and deep neural network classification models is 88.37%, 95.64% and 98.85%, respectively. Besides, for the important FPR criteria, the values 1.13, 2.56 and 10.47 are obtained for the deep neural network, C5.0 decision tree and Bayesian network, respectively. Furthermore, for the rest of the evaluation criteria, the proposed deep learning model has achieved the best results.

In addition to the above criteria, in terms of TPR, TNR, F-measure and FNR criteria, the 6-layer deep learning model performs better than the C5.0 decision tree and Bayesian network models.

Another important criterion used in this experiment is the AUC criterion, the accuracy of the surface below the ROC diagram. The better AUC value indicates the more accuracy of the model. The diagram of this criterion for the Bayesian network, C5.0 decision tree and deep learning classification models are shown in Figures 8-10, respectively.

Table 4. Results of the classification models (%).

| Classification models | ACC | TPR | TNR | PPV | F-measure | FPR | FNR |
|---|---|---|---|---|---|---|---|
| Naïve Bayes | 88.37 | 87.12 | 89.53 | 88.54 | 87.82 | 10.47 | 12.88 |
| Decision Tree of C5.0 | 95.64 | 97.3 | 97.44 | 97.15 | 95.39 | 2.56 | 2.7 |
| 6-layer Deep learning | 98.85 | 98.89 | 98.87 | 98.72 | 95.84 | 1.13 | 1.11 |

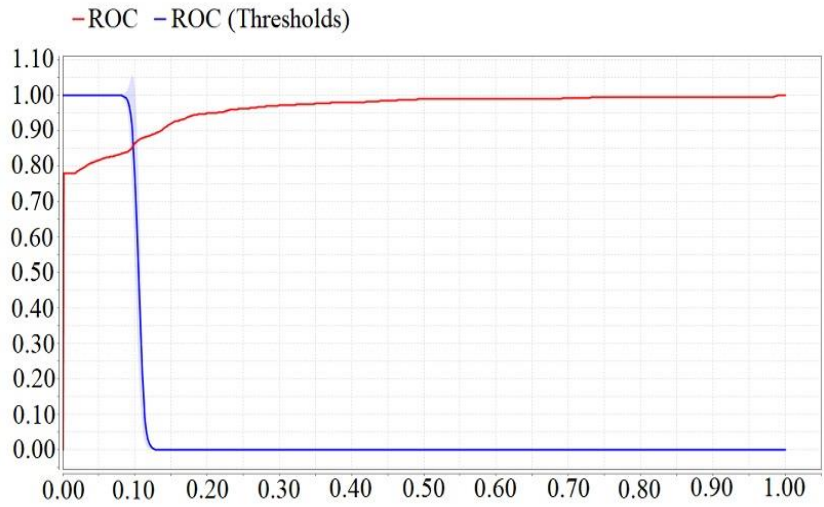

Figure 8. ROC curve for the Bayesian model.

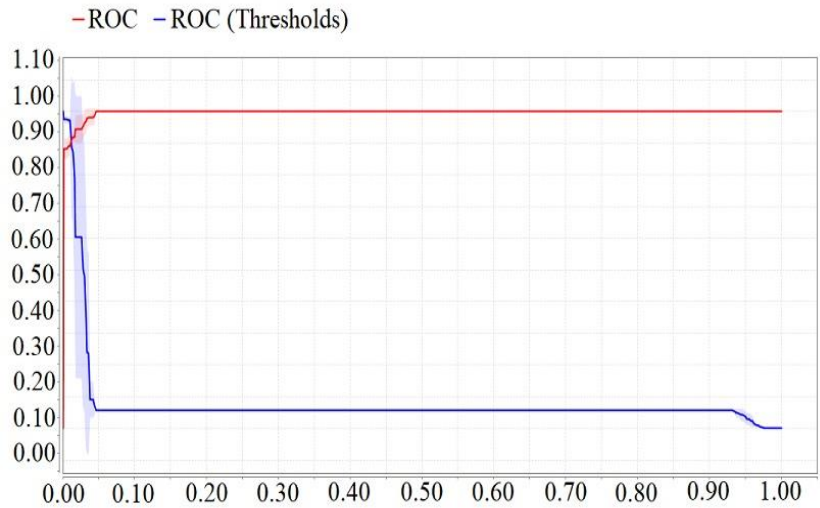

Figure 9. ROC curve for the C5.0 decision tree model.

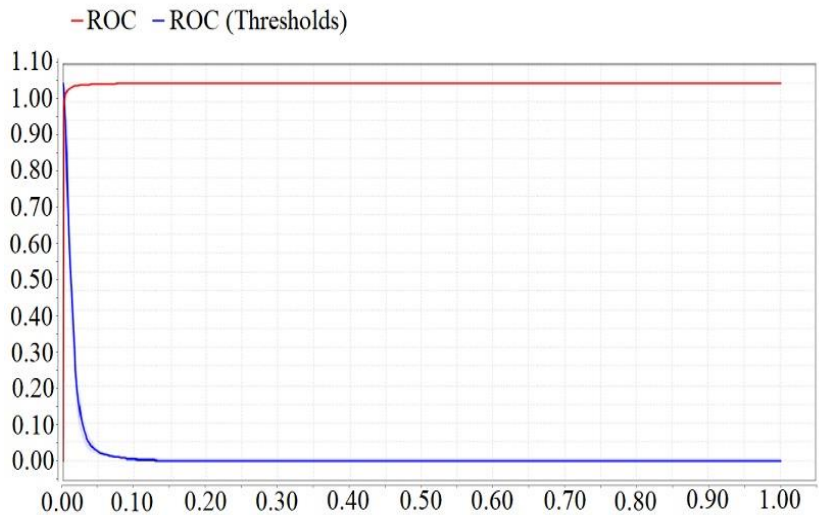

**Figure 10.** ROC curve for the 6-layer deep learning model.

According to Figures 8-10, the AUC value for the Bayesian, C5.0 decision tree and deep learning classification models are obtained as 96.1%, 99.60% and 99.90%, respectively. Consequently, the proposed deep learning model is the best model in terms of the AUC. As a result, by analyzing the classification models, it can be concluded that the 6-layer deep learning model has the best performance regarding all the criteria examined in the
output to detect APT attacks on the NSL-KDD dataset. Since the purpose of this paper is to extract features automatically in the layers related to the features using a deep neural network, the important features that are extracted using this method are explained in Figure 11.

According to Figure 11, the most important features/variables of the APT attack detection are arranged from the highest probability of occurrence to the lowest probability so that the highest probability of attack for srv_count variable, which indicates the count of connections to the identical services with the current connection during the last two seconds [35], is 1.0000.

## 6. Results and Discussions

In general, researches indicate that among the approaches developed for APT attacks detection, artificial intelligence methods are the best methods. Moreover, according to the latest scientific achievements in the field of network security, deep learning method has had the best performance comparing to other methods. Consequently, in this paper, artificial intelligence methods such as C5.0 decision tree, Bayesian network and deep neural network classification models were used to detect two normal and anomaly classes of APT attacks on the NSL-KDD dataset. The models were implemented via RapidMiner software.

As APT attack is one of the most stable and persistent attacks on the system and involves the system for a long time, it is very important to detect it early. Therefore, we needed artificial intelligence methods for timely detection of APT attacks, and we implemented three methods of C5.0 decision tree, Bayesian model and deep learning using 10-fold cross validation method.

According to Table 4, by evaluating the criteria, we concluded that the accuracy of the deep learning model reaching 98.85% is the best compared to the C5.0 decision tree and Bayesian models reaching 95.64% and 88.37%, respectively. Another important criterion is the FPR criterion, which is 1.13, 2.56, and 10.47 for the deep learning network, C5.0 decision tree and Bayesian network models, respectively. In addition, for the rest of the evaluation criteria, including TPR, TNR, F-measure and FNR, the deep learning model performed better than the C5.0 decision tree and Bayesian network models.

In addition, in terms of the AUC criterion, according to Figures 8-10, the AUC value for the Bayesian, C5.0 decision tree and deep learning models is obtained as 96.1%, 99.60% and 99.90%, respectively. Thus, regarding AUC, deep learning model is a more appropriate model for detection and classification of APT attacks.

Comparison between the C5.0 decision tree, Bayesian Network and deep learning classification models in terms of the AUC criterion via ROC curve is illustrated in Figure 12 and comparison between the models based on the mentioned criteria are shown in Figure13.

According to Figure 13, comparison between the models show that the deep learning model provides the best performance for detection and classification of APT attacks regarding ACC, TPR, TNR, PPV, F-measure, FPR, FNR and AUC criteria.

Finally, based on the results obtained, we can conclude that the deep learning model has been selected as the proposed model of this paper. As an acceptable result for the proposed deep learning model, we have implemented Lift chart [39] diagram for two normal and anomalous classes with a confidence index on the test dataset. In Figures 14 and 15, Lift chart diagram is shown for normal and anomalous classes based on the APT attack detection, respectively.

```
Variable Importances:
              Variable Relative Importance Scaled Importance Percentage
             srv_count               1.000000          1.000000   0.012395
          diff_srv_rate               0.982102          0.982102   0.012173
          service.shell               0.982063          0.982063   0.012173
        service.daytime               0.975109          0.975109   0.012086
         service.eco_i                0.972955          0.972955   0.012060
            rerror_rate               0.954541          0.954541   0.011832
     dst_host_srv_count               0.940606          0.940606   0.011659
     service.netbios_ns               0.924339          0.924339   0.011457
            service.X11               0.914025          0.914025   0.011329
         srv_rerror_rate              0.907292          0.907292   0.011246
---
        service.netstat               0.326361          0.326361   0.004045
           service.nnsp               0.325659          0.325659   0.004037
       protocol_type.tcp              0.323045          0.323045   0.004004
    service.netbios_ssn               0.321485          0.321485   0.003985
               flag.OTH               0.315218          0.315218   0.003907
                flag.S3               0.286562          0.286562   0.003552
          service.pop_2               0.257605          0.257605   0.003193
```

**Figure 11.** Features extraction using a deep neural network on the NSL-KDD dataset.

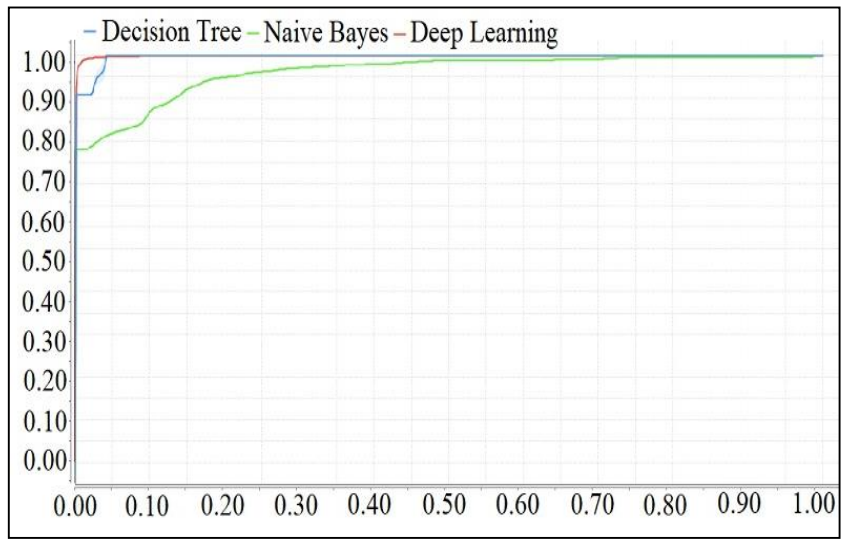

**Figure 12.** Comparison between the C5.0 decision tree, Bayesian Network and deep learning classification models in terms of the AUC criterion via ROC curve.

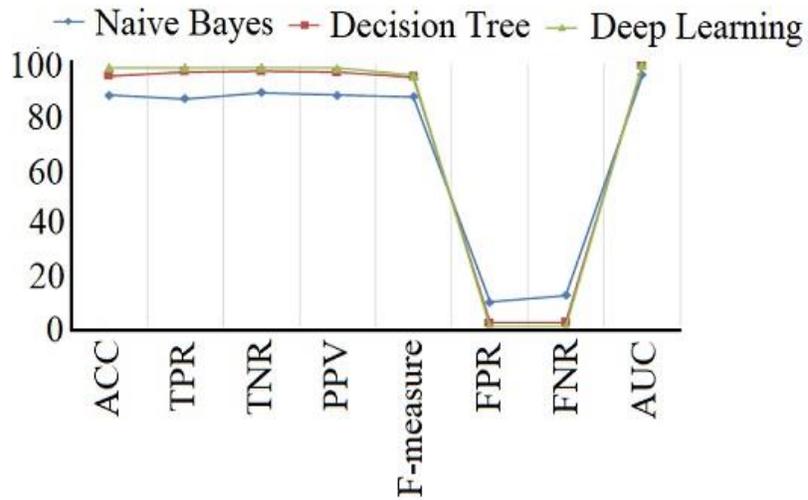

**Figure 13.** Comparison between the classification models based on the evaluation criteria.

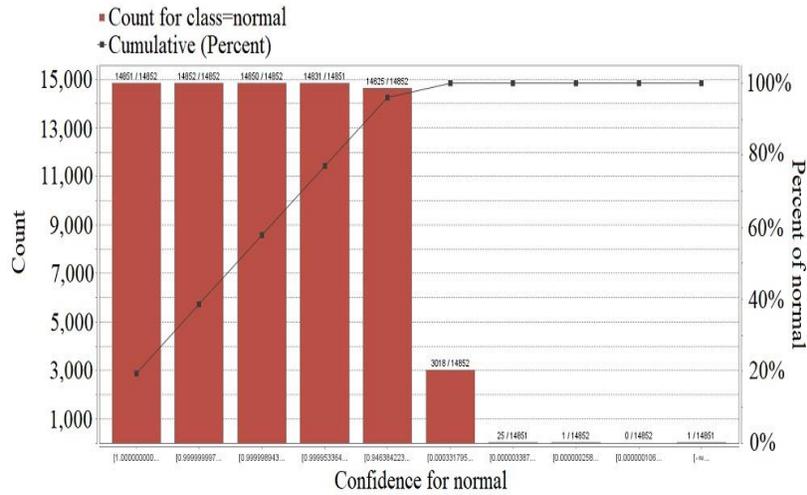

**Figure 14.** Lift Chart diagram for modeling of Deep Neural Network for normal class.

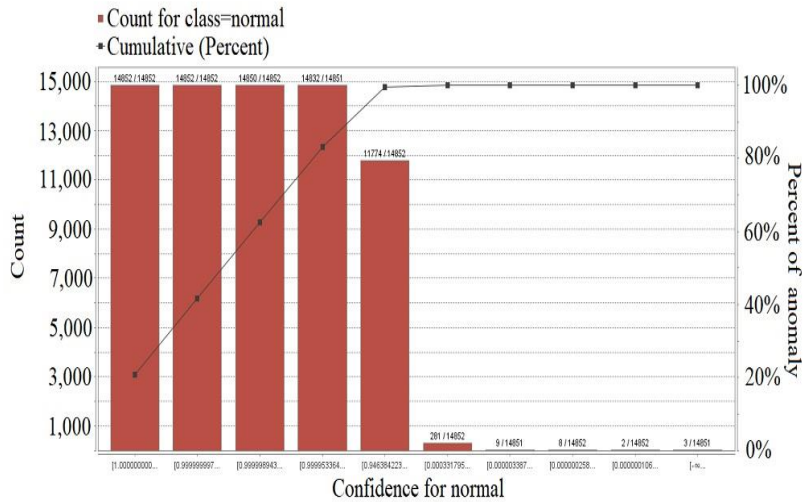

**Figure 15.** Lift Chart diagram for modeling of Deep Neural Network for anomaly class.

The diagram of Lift Chart for normal class is illustrated in Figure 14. Based on Figure 14, confidence for normal class, for example in scope 0.94 related to the fifth record include 14851 samples, illustrating that 14625 samples are as normal. Therefore, confidence for normal class at the scope 0.94 is very important for 14851 records, illustrating that over 98% of samples are as normal. Furthermore, the diagram of Lift Chart for anomaly class is shown in Figure 15. Hence, confidence for anomaly class, for example in scope 0.99 related to the fourth record contains 14851 samples illustrating that 14832 samples are as anomaly.

Therefore, confidence for anomaly class on data sets at the scope 0.99 is very important for 14851 records, illustrating that more than 98% of samples are as anomaly.

Eventually, a comparison between deep neural network model with the other works in terms of the accuracy and FPR obtained on the NSL-KDD dataset is demonstrated in Table 5.

**Table 5.** Comparison between the proposed 6-layer deep learning method and the work of other researchers in terms of the FPR and accuracy.

| ACC (%) | FPR | Technique used | Dataset used | Authors |
| --- | --- | --- | --- | --- |
| 92.84 using 40% of the training dataset | Not reported | DBN-SVM | NSL-KDD | Salama et al. [18] |
| 20.35 for U2R and 32.90 for R2L using NB, and up to 82 for DoS and 65.4 for Probe using J48 | Not reported | NB and J48 decision tree | NSL-KDD | Aziz et al. [24] |
| 81.2 for binary class and 79.9 for five classes | 3.23 for binary class and not reported for 5 classes | ANN | NSL-KDD | Ingre and Yadav [26] |
| 93.29 for KDD-99 and 95.76 for KUBD | 0.78 for KDD-99 and 1.05 for KUBD | K-NN combined with K-means | KDD-99 and KUBD | Guo et al. [28] |
| 97 | Not reported | J48 combined with AdaBoost by Information Gain | KDD-99 | Mazraeh et al. [29] |
| Not reported | Not reported | Framework | Network Traffic | Marchetti et al. [31] |
| 96.72 | 0.83 | HG-GA SVM | NSL-KDD | Raman et al. [32] |
| Not reported | Not reported | OODA LOOP | Network Traffic | Bodström and Hämäläinen, [36] |
| 84.80% | 4.5 | linear SVM | Network Traffic | Ghafir et al. [8] |
| 97.22% | Not reported | SVM-RBF | NSL-KDD | Chu et al. [9] |
| Not reported | Not reported | Deep Learning stack by exploiting sequential neural networks | Network Traffic | Bodström and Hämäläinen [14] |
| 91.80% | Not reported | Hidden Markov Model | Network Traffic | Ghafir et al. [10] |
| 98.85 | 1.13 | 6-layer deep learning model | NSL-KDD | Proposed method |

According to Table 5, we have accomplished the APT attack detection and classification process on the NSL-KDD dataset and have compared our proposed method with other researches that have used machine learning methods to detect intrusion into the NSL-KDD dataset. A comparison of the proposed 6-layer deep learning method with the work of other researchers on the NSL-KDD dataset is as follows.

Salama et al. have proposed the combined DBN-SVM method for intrusion detection, which achieved 92.84% detection accuracy and have not reported any value for the FPR [18]. In our study, the detection accuracy for APT attack is 98.85% using a 6-layer deep learning model, and the FPR value obtained via our proposed model is 1.13.

Aziz et al. have achieved 21.07% detection accuracy for R2L attack on all the training data and 32.90% accuracy on 20% of the training data using the Naive Bayes classification method [24]. Also, regarding U2R attack, 20.35% accuracy was achieved on all the training data and 16.28% accuracy was achieved on 20% of the training data. Then, they have used the J48 decision tree to achieve a high accuracy of up to 82% for DoS attack and 65.4% accuracy for Probe attack, and no FPR values were reported. Whereas, in our study, we grouped all the attack classes into a class under the anomaly attack class in the data preprocessing step. Finally, the accuracy of 98.85% is obtained for APT attack detection using a 6-layer deep learning model, and the FPR value in our proposed model is 1.13.

In a study by Ingre and Yadav, Artificial Neural Network (ANN) method has been utilized to detect intrusion [26]. Their results show that the ANN method with detection accuracies of 81.2% and 79.9% for 2 classes and 5 classes, respectively, has better performance comparing to the SOM method with 75.49% detection accuracy, and also, the FPR obtained via their method is 3.23 for the binary class. While, in our study, the accuracy of 98.85% is achieved using a 6-layer deep learning model, and the FPR value is obtained as 1.13.

In the next study, Raman et al. have proposed an intrusion detection technique using a Hyper Graph-Based Genetic Algorithm (HG-GA) for parameterization and feature selection in the Support Vector Machine (SVM) [32]. They achieved 96.72% accuracy using the proposed HG-GA SVM method, and also the FPR value obtained applying their method is 0.83. But in our study, the accuracy of APT attack detection using a 6-layer deep learning model is 98.85%, and the FPR value is 1.13.

Chu et al. have used the combined SVM-RBF method to detect APT attacks with 97.22% accuracy and the FPR value has not been reported in their work [9]. However, we achieved the accuracy of 98.85% using a 6-layer deep learning model, and the FPR value through our proposed model is 1.13.

As a general result, the proposed deep learning model with the extracted features according to Figure 11 has the best performance in comparison with the work of others in terms of the above evaluation criteria for APT attack detection. To the best of our knowledge, it is the first time that the attack classes are grouped into an anomaly class to detect an APT anomaly attack on the NSL-KDD dataset in the data preprocessing step. Moreover, for the first time, the 10-fold cross validation method is utilized for APT attack detection, so that all the training and testing data are aggregated and integrated, and then, the 10-fold cross validation method is applied.

## 7. Conclusion and Future Works

In this study, three artificial intelligence-based classification models including Bayesian Network, C5.0 decision tree and deep learning were used to detect and classify APT attacks on the NSL-KDD dataset. Since the nature of the APT attack is permanent and persistent presence in the victim system, early detection of this attack requires high accuracy and minimal FPR in the early steps. For this purpose, through the mentioned classification models, based on the obtained results, a 6-layer deep learning model with the highest accuracy and the lowest FPR, which are equal to 98.85 and 1.13, respectively, was selected as the final model. In addition, other evaluation criteria, such as TPR, TNR, PPV, F-measure, FPR, FNR and AUC were investigated. The 6-layer deep learning model had also the best performance in terms of these criteria. One of the important criteria for comparing models is the AUC criterion. Figures 8-10 as well as Figure 12, comparing the three classification models, show that the deep learning model with the AUC value 99.9% is better than the Bayesian Network and C5.0 decision tree models with the AUC values 99.6% and 99.60%, respectively.

Finally, Table 5 summarizes the comparison of the proposed deep learning model with other related works on the NSL-KDD dataset. The 6-layer deep learning model had the best execution and performance in terms of the accuracy compared to previous work [9] regarding APT attack detection on the NSL-KDD dataset. Furthermore, so far in no study the important features of the dataset have been extracted. Figure 11 shows the importance of the features. As an important result, deep learning has been ranked the highest and best in most areas of network security detection, and in this article, we also have obtained the best results for the deep learning model. For future work, we suggest that a combination of machine learning and deep learning methods can be implemented on the NSL-KDD dataset used and network traffic flow. Moreover, supervised and unsupervised deep learning methods, such as Recurrent Neural Networks and Auto-Encoder Neural Networks, respectively, can be utilized.

## References

[1]. Y. Wang, Q. Li, Z. Chen, P. Zhang, and G. Zhang, "A Survey of Exploitation Techniques and Defenses for Program Data Attacks," *J. Netw. Comput. Appl.*, vol. 154, p. 102534, 2020.

[2]. J. Chen, C. Su, K.-H. Yeh, and M. Yung, "Special issue on advanced persistent threat." Elsevier, 2018.

[3]. S. Singh, P. K. Sharma, S. Y. Moon, D. Moon, and J. H. Park, "A comprehensive study on APT attacks and countermeasures for future networks and communications: challenges and solutions," *J. Supercomput.*, vol. 75, no. 8, pp. 4543–4574, 2019.

[4]. A. Alshamrani, S. Myneni, A. Chowdhary, and D. Huang, "A survey on advanced persistent threats: Techniques, solutions, challenges, and research opportunities," *IEEE Commun. Surv. Tutorials*, vol. 21, no. 2, pp. 1851–1877, 2019.

[5]. M. Auty, "Anatomy of an advanced persistent threat," *Netw. Secur.*, vol. 2015, no. 4, pp. 13–16, 2015.

[6]. https://malpedia.caad.fkie.fraunhofer.de/actor/turla_group

[7]. https://www.cynet.com/cyber-attacks/advanced-persistent-threat-apt-attacks/

[8]. I. Ghafir *et al.*, "Detection of advanced persistent threat using machine-learning correlation analysis," *Futur. Gener. Comput. Syst.*, vol. 89, pp. 349–359, 2018.

[9]. W.-L. Chu, C.-J. Lin, and K.-N. Chang, "Detection and Classification of Advanced Persistent Threats and Attacks Using the Support Vector Machine," *Appl. Sci.*, vol. 9, no. 21, p. 4579, 2019.

[10]. I. Ghafir *et al.*, "Hidden Markov models and alert correlations for the prediction of advanced persistent threats," *IEEE Access*, vol. 7, pp. 99508–99520, 2019.

[11]. M. Lee and D. Lewis, "Clustering disparate attacks: Mapping the activities of the advanced persistent threat," *Last accessed June*, vol. 26, 2013.

[12]. B. Bencsáth, G. Pék, L. Buttyán, and M. Félegyházi, "Duqu: Analysis, detection, and lessons learned," in *ACM European Workshop on System Security (EuroSec)*, 2012, vol. 2012.advanced persistent threat," *Last accessed June*, vol. 26, 2013.

[13]. M. Balduzzi, V. Ciangaglini, and R. McArdle, "Targeted attacks detection with spunge," in *2013 Eleventh Annual Conference on Privacy, Security and Trust*, 2013, pp. 185–194.

[14]. T. Bodström and T. Hämäläinen, "A novel deep learning stack for APT detection," *Appl. Sci.*, vol. 9, no. 6, p. 1055, 2019.

[15]. B. Mukherjee, L. T. Heberlein, and K. N. Levitt, "Network intrusion detection," *IEEE Netw.*, vol. 8, no. 3, pp. 26–41, 1994.

[16]. M. Roesch, "Snort: Lightweight intrusion detection for networks.," in *Lisa*, 1999, vol. 99, no. 1, pp. 229–238.

[17]. D. E. Denning, "An intrusion-detection model," *IEEE Trans. Softw. Eng.*, no. 2, pp. 222–232, 1987.

[18]. M. A. Salama, H. F. Eid, R. A. Ramadan, A. Darwish, and A. E. Hassanien, "Hybrid intelligent intrusion detection scheme," in *Soft computing in industrial applications*, Springer, 2011, pp. 293–303.

[19]. R. Brewer, "Advanced persistent threats: minimising the damage," *Netw. Secur.*, vol. 2014, no. 4, pp. 5–9, 2014.

[20]. N. Virvilis and D. Gritzalis, "The big four-what we did wrong in advanced persistent threat detection?," in *2013 international conference on availability, reliability and security*, 2013, pp. 248–254.

[21]. B. Bencsáth, G. Pék, L. Buttyán, and M. Felegyhazi, "The cousins of stuxnet: Duqu, flame, and gauss," *Futur. Internet*, vol. 4, no. 4, pp. 971–1003, 2012.

[22]. T. M. Chen and S. Abu-Nimeh, "Lessons from stuxnet," *Computer (Long. Beach. Calif).*, vol. 44, no. 4, pp. 91–93, 2011.

[23]. M. H. Au, K. Liang, J. K. Liu, R. Lu, and J. Ning, "Privacy-preserving personal data operation on mobile cloud—Chances and challenges over advanced persistent threat," *Futur. Gener. Comput. Syst.*, vol. 79, pp. 337–349, 2018.

[24]. A. S. A. Aziz, A. E. Hassanien, S. E.-O. Hanaf, and M. F. Tolba, "Multi-layer hybrid machine learning techniques for anomalies detection and classification approach," in *13th International Conference on Hybrid Intelligent Systems (HIS 2013)*, 2013, pp. 215–220.

[25]. J. R. Johnson and E. A. Hogan, "A graph analytic metric for mitigating advanced persistent threat," in *2013 IEEE International Conference on Intelligence and Security Informatics*, 2013, pp. 129–133.

[26]. B. Ingre and A. Yadav, "Performance analysis of NSL-KDD dataset using ANN," in *2015 international conference on signal processing and communication engineering systems*, 2015, pp. 92–96.

[27]. I. Friedberg, F. Skopik, G. Settanni, and R. Fiedler, "Combating advanced persistent threats: From network event correlation to incident detection," *Comput. Secur.*, vol. 48, pp. 35–57, 2015.

[28]. C. Guo, Y. Ping, N. Liu, and S.-S. Luo, "A two-level hybrid approach for intrusion detection," *Neurocomputing*, vol. 214, pp. 391–400, 2016.

[29]. S. Mazraeh, M. Ghanavati, and S. H. N. Neysi, "Intrusion detection system with decision tree and combine method algorithm," *Int. Acad. J. Sci. Eng.*, vol. 3, no. 8, pp. 21–31, 2016.
37